\documentclass[sigconf]{acmart}
\AtBeginDocument{%
  }

\setcopyright{none}
\copyrightyear{2025}
\acmYear{2025}
\acmDOI{XXXXXXX.XXXXXXX}
\acmConference[Preprint]{...(Accepted for publication at ICSE 2026)}{}{final version to appear in 2026}
\usepackage{todonotes}
\usepackage{lipsum}
\usepackage{listings}

\usepackage{booktabs}
\usepackage{tabularx}
\usepackage{soul}
\usepackage{graphicx}
\usepackage{subcaption}
\usepackage{multirow}
\usepackage[table,xcdraw]{xcolor}

\newcommand{\qwen}{qwen2.5-coder:32b}
\newcommand{\llama}{llama3:70b}

\lstdefinelanguage{json}{
    basicstyle=\ttfamily\small, % Use a fixed-width font
    numbers=left, % Line numbers on the left
    numberstyle=\tiny\color{gray}, % Style of line numbers
    stepnumber=1, % Increment of line numbers
    showstringspaces=false, % Do not show spaces in strings
    breaklines=true, % Allow line breaking
    frame=single, % Frame around the listing
    backgroundcolor=\color{lightgray!20}
    , % Light gray background
    literate=
     *
      {:}{{{\color{red}:}}}{1}%
      {,}{{{\color{red},}}}{1}%
      {\{}{{{\color{orange}\{}}}{1}%
      {\}}{{{\color{orange}\}}}}{1}%
      {[}{{{\color{orange}[}}}{1}%
      {]}{{{\color{orange}]}}}{1}%
}

\begin{document}

%%
%% The "title" command has an optional parameter,
%% allowing the author to define a "short title" to be used in page headers.
%\title{LLMs can make system interoperate autonomously} NO! We cannot guarantee it
%\title{Evaluating LLM-based interoperability}
\title{Evaluating the effectiveness of LLM-based interoperability}
%\title{Some LLMs can make system interoperate autonomously}  NO! We cannot guarantee it
%\title{Can open-source LLMs make system interoperate autonomously?}
%\title{Can LLMs make system interoperate autonomously?}

%%
%% The "author" command and its associated commands are used to define
%% the authors and their affiliations.
%% Of note is the shared affiliation of the first two authors, and the
%% "authornote" and "authornotemark" commands
%% used to denote shared contribution to the research.
\author{Rodrigo Falc\~{a}o}
\email{rodrigo.falcao@iese.fraunhofer.de}
\orcid{0000-0003-1222-0046}
%\email{rodrigo.falcao@fraunhofer.de}
\affiliation{%
  \institution{Fraunhofer IESE}
  \city{Kaiserslautern}
 % \state{Ohio}
  \country{Germany}
}

\author{Stefan Schweitzer}
\affiliation{%
  \institution{Fraunhofer IESE}
  \city{Kaiserslautern}
 % \state{Ohio}
  \country{Germany}
}
%\email{stefan.schweitzer@iese.fraunhofer.de}

\author{Julien Siebert}
\orcid{0000-0002-7696-0046}
\affiliation{%
  \institution{Fraunhofer IESE}
  \city{Kaiserslautern}
 % \state{Ohio}
  \country{Germany}
}
%\email{stefan.schweitzer@iese.fraunhofer.de}

\author{Emily Calvet}
\affiliation{%
  \institution{Fraunhofer IESE}
  \city{Kaiserslautern}
 % \state{Ohio}
  \country{Germany}
}
%\email{frank.elberzhager@iese.fraunhofer.de}
\orcid{0000-0003-4374-9129}
\authornote{Currently affiliated to SAP.}

\author{Frank Elberzhager}
\affiliation{%
  \institution{Fraunhofer IESE}
  \city{Kaiserslautern}
 % \state{Ohio}
  \country{Germany}
}
%\email{frank.elberzhager@iese.fraunhofer.de}
\orcid{0000-0002-8748-3927}

%%
%% By default, the full list of authors will be used in the page
%% headers. Often, this list is too long, and will overlap
%% other information printed in the page headers. This command allows
%% the author to define a more concise list
%% of authors' names for this purpose.
\renewcommand{\shortauthors}{Falc\~{a}o et al.}

%%
%% The abstract is a short summary of the work to be presented in the
%% article.
\begin{abstract}
  \textit{Background:} Systems of systems are becoming increasingly dynamic and heterogeneous, and this adds pressure on the long-standing challenge of interoperability. Besides its technical aspect, interoperability has also an economic side, as development time efforts are required to build the interoperability artifacts.
  \textit{Objectives:} With the recent advances in the field of large language models (LLMs), we aim at analyzing the effectiveness of LLM-based strategies to make systems interoperate autonomously, at runtime, without human intervention.
  \textit{Method:} We selected 13 open source LLMs and curated four versions of a dataset in the agricultural interoperability use case. We performed three runs of each model with each version of the dataset, using two different strategies. Then we compared the effectiveness of the models and the consistency of their results across multiple runs.
  \textit{Results:} \qwen~ was the most effective model using both strategies DIRECT (average pass@1 $\geq 0.99$) and CODEGEN (average pass@1 $\geq 0.89$) in three out of four dataset versions. In the fourth dataset version, which included an unit conversion, all models using the strategy DIRECT failed, whereas using CODEGEN \qwen~ succeeded with an average pass@1 = 0.75.
  \textit{Conclusion:} Some LLMs can make systems interoperate autonomously. Further evaluation in different domains is recommended, and further research on reliability strategies should be conducted.
\end{abstract}

%%
%% The code below is generated by the tool at http://dl.acm.org/ccs.cfm.
%% Please copy and paste the code instead of the example below.
%%


%%
%% Keywords. The author(s) should pick words that accurately describe
%% the work being presented. Separate the keywords with commas.
%% A "teaser" image appears between the author and affiliation
%% information and the body of the document, and typically spans the
%% page.
% \begin{teaserfigure}
%   \includegraphics[width=\textwidth]{sampleteaser}
%   \caption{Seattle Mariners at Spring Training, 2010.}
%   \Description{Enjoying the baseball game from the third-base
%   seats. Ichiro Suzuki preparing to bat.}
%   \label{fig:teaser}
% \end{teaserfigure}

\received{20 February 2007}
\received[revised]{12 March 2009}
\received[accepted]{5 June 2009}

%%
%% This command processes the author and affiliation and title
%% information and builds the first part of the formatted document.
\maketitle

\section{Motivation}\label{sec:intro}

We have experienced an age of unpaired ability to collect data and network systems, and also a trend toward an increasingly connected world, where systems become systems of systems, which in turn are growing in scale, complexity, and heterogeneity to give rise to what has been referred to as dynamic software ecosystems \cite{pelliccione2023architecting} or dynamic systems of systems \cite{adler2024defining}.
%Pelliccione et al. \cite{pelliccione2023architecting} elaborate on the same type of system as dynamic software ecosystems.
These systems are characterized, among other things, by their openness and heterogeneity. This implies that the constituent systems of a dynamic system of systems cannot be known in advance, and interoperability needs to be achieved via black-box integration.

%Yet, interoperability is a major challenge in black-box integration, and one of the reasons is that it is difficulty to achieve a common understanding of the data formats, standards, and interfaces among different stakeholders \cite{liu2020human}.

\subsection{Problem}\label{sec:problem-statement}
Interoperability is the “capability of a product to exchange information with other products and mutually use the information that has been exchanged” \cite{iso25010-2023} and is recognized as a long-standing challenge in virtually all software-intensive systems across multiple domains \cite{maciel2024systems}. The challenge lingers not because it has never been solved, especially concerning technology\footnote{While there have been several attempts to classify interoperability types, when it comes to the technological aspects of interoperability, the most stable type definitions are technical, syntactic, and semantic. For a comprehensive and up-to-date survey on interoperability types, see \cite{maciel2024systems}.}: Systems can interoperate using various well-defined protocols, which solves the data exchange problem at the technical level. There are also several standard data formats serving the most diverse purposes that help address the syntactic level of interoperability. Standards for adding semantics to the data, such as semantic annotations, have been available for decades.

Notwithstanding, interoperability remains a persistent challenge, for at least three reasons. The first is that new technologies appear over time, bringing up new needs regarding how digital products must exchange information \cite{maciel2024systems}. The second is organizational: interoperability is a major challenge in black-box integration, and one of the reasons is that it is difficult to achieve a common understanding of the data formats, standards, and interfaces among different stakeholders \cite{liu2020human}. The third is rather economical, for achieving interoperability comes with a cost: engineers must implement the necessary artifacts for interoperability at design time \cite{maciel2024systems}, and these are specific to each pair of systems. This has also been observed by Stegemann and Gersch \cite{stegemann2019interoperability} in the digital health context. The authors ``emphasize that the problem of lack of interoperability is less related to insufficient technical standards as it is to economic issues'', where the direct costs of interoperability refer to the data adaptation process. Each individual system participating in a data exchange has its own representation of the domain, which fits their organizational communication structure (see Conway’s law \cite{conway1968committees}). For example, consider two systems exchanging data through an API. Even if they adopt de facto standards for implementing their APIs (e.g., REST APIs over HTTP with JSON as data format), understanding the structure of the data (syntax) and its meaning (semantics) requires human-based efforts for reading and understanding API documentation to implement, test, and deploy the software artifacts that are actually responsible for exchanging data and making it usable. Technological solutions such as the Semantic Web, which aimed at enabling systems to interoperate without knowing anything about the data at design time \cite{lassila2001semantic}, despite being available for decades, never took off completely -- recurrent criticisms include adoption costs, being a niche technology, and the risk of becoming redundant with AI advancements, among others \cite{hogan2020semantic}.

\subsection{Research question}

In recent years, large language models (LLMs) have emerged as a tool with considerable potential to address problems in numerous fields. In essence, an LLM is a probabilistic model trained on extensive data to generate meaningful word sequences \cite{dhar2024can}. One popular application field of LLMs has been software development, where several coding assistants have been proposed (e.g., GitHub Copilot\footnote{\url{https://docs.github.com/en/copilot/about-github-copilot}}, OpenAI Codex\footnote{\url{https://openai.com/index/openai-codex/}}). These solutions are based on models that have been particularly trained to support coding tasks.

These advancements have led us to consider different ways to address the interoperability challenge. What if we could shift to runtime the design time efforts associated with understanding data representations, implementing data adapters, and deploying interoperability solutions? What if systems were ``smart enough'' to exchange and use data without human intervention? More precisely, our request question is: \textit{Can LLMs make systems interoperate autonomously?} To answer this question, in this paper, we contribute (1) an empirical evaluation of two potential LLM-based strategies for achieving interoperability and (2) four versions of a manually curated dataset containing data in different representations in an agricultural use case. The results indicate that some models can be highly effective at making systems interoperate autonomously.

To structure this paper, we have adapted the guidelines of Jedlitschka and Pfahl \cite{jedlitschka2005reporting}. As to the LLM-specific aspects, we followed the guidelines\footnote{Also available at \url{https://llm-guidelines.org}.} proposed by Wagner et al \cite{wagner2025towards}. The remainder of this paper is organized as follows: Section~\ref{sec:related-work} presents related work; Section~\ref{sec:strategies} described the two LLM-based strategies; Section~\ref{sec:method} details the research method; Section~\ref{sec:results} reports the evaluation results; Section~\ref{sec:interpretation} discusses the findings and their limitations; and Section~\ref{sec:conclusion} concludes the article.

\section{Related work}\label{sec:related-work}

Interoperability is a vast area with multiple levels, and numerous studies over the decades have explored ways to address it. Here we focus on recent works that incorporate LLMs into their approaches.

Yuan et al. \cite{yuan2023large} aim to shorten the gap of interoperability between Electronic Health Records (EHRs) and clinical trial criteria. The authors suggest enhancing patient-trial matching by augmenting healthcare data using an LLM-based method. This means matching patients with appropriate clinical trials with the aid of skilled computer tools. To protect private patient data and still take advantage of LLM capabilities, a privacy-aware data augmentation strategy was presented. To close the gap between clinical trial criteria and EHRs, the authors created a method known as LLM-based patient-trial matching (LLM-PTM). This was done to improve compatibility and accelerate the matching process. To determine whether the LLM-PTM approach is effective, the researchers ran tests and utilized known evaluation metrics, such as precision, recall, and F1 score. The evaluation also examines the model's performance across different trials, highlighting its consistent and robust capacity to handle challenging tasks compared to the baseline model. However, it was not specified how exactly LLMs were used in the implementation or if it is intended for runtime execution, and it does not improve the interoperability of similar systems but of clinical trials with EHRs.

Expanding on the healthcare domain, Li et al. \cite{li2024fhir} aim to improve the exchange of health data across diverse platforms by harnessing LLMs to generate FHIR-formatted resources from free-text input. They have developed a model called FHIR-GPT model for the format transformation, and compared the results with other three LLMs (OpenAI GPT-4, Llama-2-70B, and Falcon-180B). %which was performed by prompting the model with the task instruction, expected output template in JSON format, 4-5 conversion examples, a list of medical codes to be chosen from, and, finally, the input text. 
They employed a few-shot approach, providing the models with examples of the desired transformations, to make the model output the transformed data. The model was tested with a subset of 100 dataset entries, and then the discrepancies between the LLM-generated FHIR output and their human annotations were manually reviewed to adjust the prompts.
%The testing was performed through multiple asynchronous API calls. 
Their primary criteria of evaluation was the exact match rate with the human annotations previously converted to FHIR and validated to be used as a gold standard. The FHIR-GPT outperformed the other models, achieving a 90\% accuracy rate and an F1 score greater than 0.96. They found the accuracy of the output to be highly dependent on the prompts used and brought attention to some recommendations to be considered.% This study did not mention if the solution is envisioned to be used at runtime or development time.

Santos et al. \cite{santos2025interactive} introduce an LLM-based agentic system to help users harmonize datasets through interactions with the users. Their system, named Harmonica, provides a user interface where users can provide a dataset, check the data harmonization suggestions, and improve it by interacting with the agent. The LLM agent is placed between the user and a series of ``data integration primitives'', a set of programs to solve well-defined data integration tasks. The tool has been demonstrated in a health-related use case, mapping clinical data to the GDC standard. As human interaction is required, the actual data harmonization is not fully autonomous, and LLMs play only a direct role in converting data by interfacing the user with the data integration primitives.

Berenguer et al. \cite{Berenguer2024UsingLLMs} explores how LLMs might improve the reusability of sensor data by bridging the semantic gaps between different data formats. The study goes into great detail about the problems with sensor data interoperability, emphasizing how these problems prevent sensor data from being accessible and reusable across systems. It proposes the use of LLMs to convert sensor data initially presented in non-interoperable formats like HTML into more usable formats like JSON or XML. The authors mention that the latency, response time, and scalability are crucial factors to be considered when dealing with application that require quick results. The study evaluates the LLMs, specifically GPT-4, GPT-3.5, and Llama 2, in translating HTML to JSON/XML. GPT-4 showed a high recall rate of 85.33\% and a precision rate of 93.51\%. The study used data obtained from 25 different sensors and did a manual qualitative analysis of a randomly selected sample of 10 sensors, where it was noted that the models can eventually hallucinate and rename some values. The authors do not explicitly mention that their solution is meant to used at runtime, but their considerations to scalability and response time makes us gravitate to that conclusion.

Lehmann \cite{lehmann2024towards} elaborates on the potential of using LLMs to make different APIs interoperate. The author proposes a middleware composed of two ``translators'', one on each side of the interaction between an API provider and an API consumer. These translators should be LLM-based and trained on knowledge about the application they are attached to. When the API consumer wants to call an API of the API provider, it should perform a function call two its translator, which in turn would call the translator on the side of the provider using natural language. The provider's translator should transform the natural language call into the format of the provider call (i.e., function transformation). When the provider returns the data, its translator sends it back to the consumer translator in natural language, which returns the data to the consumer as need (i.e., data transformation). In the current stage, the idea has not been implemented nor evaluated.

Abu-Salih et a. \cite{abu2025using} conducted a systematic literature review on using LLMs for semantic interoperability (SI), identifying schema/ontology alignment and dynamic interoperability among the primary applications. For the alignment application, Xia et al. \cite{xia2024generation} use LLMs to automate the generation of I4.0 digital twins. Regarding dynamic interoperability, He at el. \cite{he2023exploring} is mentioned as an example of an idea of using a zero-shot setting for ontology matching, which has been evaluated one model (Flan-T5-XXL) processing two datasets and measured precision, recall, and F1-score.

While other works are either limited in scope concerning the number of LLMs evaluated, not fully autonomous (i.e., human in the loop is needed), or still preliminary, in this paper we report on the evaluation of two fully autonomous LLM-based strategies to make systems interoperate using several models, measuring and comparing effectiveness and consistency of LLMs in implementing the strategies. To the best of our knowledge, there are no similar empirical studies in the literature.

\section{The two strategies}\label{sec:strategies}

Consider the following scenario: a software-based service S needs to get data from external services, which each might deliver data in different representations (i.e., formats and schemas). In a traditional approach, at least one participant in this data exchange must understand the other system's data representation, implement a data adapter, test, and deploy the solution before the two systems can interact. Instead, we propose that the service S should be able to process data in any arbitrary unknown representation and convert it into its internal representation, fully autonomously (i.e., no human in the loop) and on the fly.  To investigate this idea, we elaborated and implemented two strategies for the service S using LLMs.

\begin{figure*}[htbp]
    \centering
    % First subfigure
    \begin{subfigure}[b]{0.35\textwidth}
        \centering
        \includegraphics[width=\textwidth]{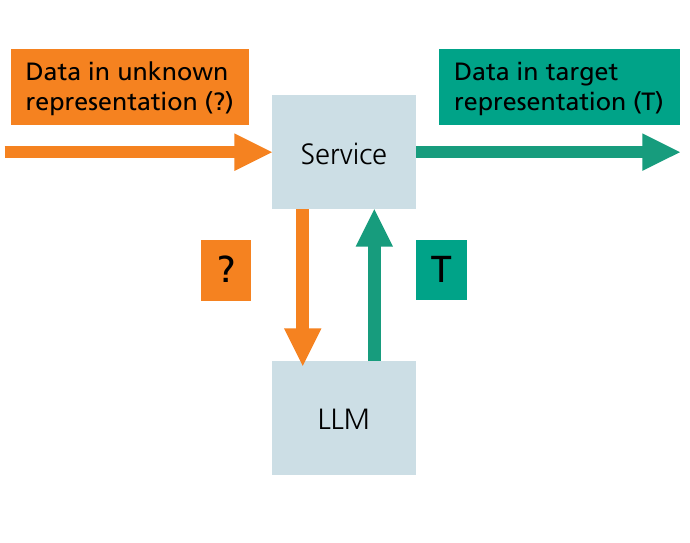} % Replace with your image file
        \caption{\textit{Direct conversion}: the input data comes into the service with an unknown representation (?), and the service prompts the LLM to convert the data into the desired target representation (T).}
        \label{fig:direct-conversion}
    \end{subfigure}
    \hfill
    % Second subfigure
    \begin{subfigure}[b]{0.55\textwidth}
        \centering
        \includegraphics[width=\textwidth]{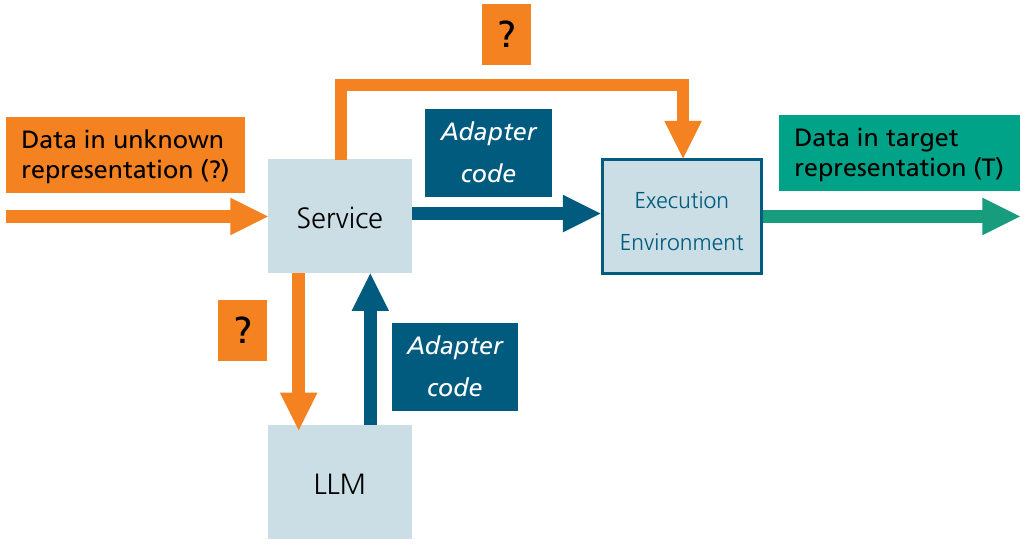} % Replace with your image file
        \caption{\textit{Conversion module generation}: the input date comes into the service with an unknown representation (?), and the service prompts the LLM to create an algorithm to convert the input data into the desired target representation (T). Then, the service deploys the generated code and calls it providing \textit{?} as input.}
        \label{fig:codegen}
    \end{subfigure}

    \caption{Overview of the two implemented strategies for LLM-based interoperability.}
    \label{fig:two-strategies}
\end{figure*}

\subsection{Direct conversion}\label{sec:direct-conversion}
We name the first strategy \textit{direct conversion} (DIRECT), where the LLMs are prompted to produce the adapted version of a given input data directly. Figure~\ref{fig:direct-conversion} illustrates this strategy, which consists of relying on the potential ability of the models to ``understand'' the syntactics and semantics of the input data representation (which is unknown) and its relation with the target data representation (which is known). Imagine that the service S has a JSON-based representation for the entity Person. The service could build a prompt dynamically from a template with a placeholder for the input data, as follows\footnote{This is a simplification for illustration purposes only. The actual prompts used in our evaluation are provided in the experimentation package.}:

\begin{verbatim}
Consider the data below:
```
<PLACEHOLDER-FOR-INPUT>
```
Convert it into the following representation:
```
{
  "person": {
    "firstName": "John",
    "familyName": "Doe"
  }
}
```

\end{verbatim}

Now, consider that an external system has Person data in XML as follows and sends it to the service S:

\begin{verbatim}
<?xml version="1.0" encoding="UTF-8"?>
 <person>
  <name>Alice J. Smith</name>
  <address>221b Baker St, London NW1 6XE, UK </adress>
 </person>
</xml>
\end{verbatim}

If the model works as desired, such a prompt should output a JSON version of the input XML data that follows the schema provided in the example that has been included in the prompt:

\begin{verbatim}
{
  "person": {
    "firstName": "Alice",
    "familyName": "Smith"
  }
}
\end{verbatim}

\subsection{Conversion module generation}

In the strategy \textit{conversion module generation} (CODEGEN), the service prompts the LLMs to create an algorithm that can convert the provided input data (unknown representation) into its internal (i.e., known) representation. Figure~\ref{fig:codegen} illustrate this strategy. The LLM generates source code that contains a function that receives the unknown input data as a parameter and converts it into the desired target representation. The code is cleaned, tested, and deployed in an execution environment, where it receives the unknown input data and outputs the converted data in the target representation. Once available, the generated code can be reused. Similar to what has been shown in the previous strategy, the service S builds a prompt dynamically using a prompt template with a placeholder for the input data.

\section{Experimental plan}\label{sec:method}

In this section, the experimental plan is described. All elements here mentioned, including datasets, programs, and analysis scripts, as well as the results of the experiment (see Section~\ref{sec:results}) are available in the experimentation package\footnote{\url{https://doi.org/10.5281/zenodo.15913264}}.

\subsection{Goal, questions, and metrics}

We have formalized our research goal using the GQM template \cite{basili1994goal} as follows: \textit{To analyze the usage of the LLM-based interoperability strategies ``DIRECT'' and ``CODEGEN'' for the purpose of evaluation with respect to their effectiveness from the point of view of the researchers in a controlled setting.} To achieve the goal, we derived the questions and corresponding metrics listed in Table~\ref{tab:questions-metrics}.

\begin{table}[h]
  \caption{Questions and metrics.}
  \label{tab:questions-metrics}
\begin{tabularx}{0.5\textwidth}{>{\raggedright\arraybackslash}X>{\raggedright\arraybackslash}X}
    \hline
    \textbf{Question} & \textbf{Metric} \\
    \hline
    \textbf{Q1:} How effective are the LLM-based interoperability strategies? & \textbf{M1.1:} Average effectiveness after N runs (for each strategy and model); \textbf{M1.2:} Comparison of average effectiveness measurements. \\
    \hline
    \textbf{Q2:} Is the effectiveness of the LLM-based interoperability strategies consistent across multiple runs? & \textbf{M2:} Comparison of the effectiveness of each run (for each strategy and model) \\
    \hline
    \textbf{Q3:} Does the effectiveness of the LLM-based interoperability strategies vary depending on the dataset version? & \textbf{M3:} Comparison of the average effectiveness using different dataset versions (for each strategy and model)  \\
    \hline
    \textbf{Q4:} What are the most common failure causes of the LLM-based interoperability strategies? & \textbf{M4:} Number of failure types per strategy \\
    \hline
 
    \hline

\end{tabularx}
\end{table}

To measure the effectiveness of the models, we used pass@k, an established metric to evaluate the functional correctness of generative models trained on code \cite{chen2021evaluating}. In this evaluation, we have set $k=1$, for pass@1 is adequate to evaluate zero-shot approaches. The reason is that we are targeting use cases where the system should be able to convert data at runtime autonomously, i.e., without previous knowledge about the input representation. Therefore, a few-shot approach (which implies previous knowledge and the provision of examples) would not be suitable. 
As pass@1, numerically, is a proportion of the correct top-1 predictions of a model for entries of a given dataset, we applied the two proportion z-test \cite{bobbitt2020ztest} to compare the effectiveness of different models (M2) and the effectiveness of each run (M3).

\subsection{Use case}

We aimed for a use case that, at the same time, reflected a production interoperability scenario and was not too trivial. For a production use case, we mean data formats and schemas that are used in practice; as for triviality, we refer to scenarios where the data schema is too simple (as in the example that illustrates Section~\ref{sec:direct-conversion}).

Based on these criteria, we have selected a use case from the agricultural domain. In agriculture, field data is a core asset, including the field name, its geographic boundaries, the crop type, the crop maturity, and the soil humidity level, among various other attributes \cite{falcao2023reference}. Field boundaries, specifically, are key information in agriculture; there are many representations with no commonly agreed standard. More concretely, in our experiment, we use LLMs to convert field boundaries from a representation based on the proprietary schema of John Deere's API\footnote{See John Deere’s API documentation for field boundaries at \url{https://developer.deere.com/dev-docs/boundaries} (visited on Aug 15 2024).} to a target representation in GeoJSON \cite{rfc7946}.

\subsection{Dataset preparation}\label{sec:dataset-prep}

We created a dataset containing 222 field boundaries from real agricultural fields. The dataset is organized as a directory with three sets of files: 222 input files (based on John Deere's representation), 222 expected output files (GeoJSON), and one target file (an example of the expected output file in GeoJSON). First, we collected manually field boundaries of real agricultural fields in GeoJSON using geojson.io\footnote{\url{https://geojson.io}}. Then we created a conversion script to convert the GeoJSON field boundaries into the John Deere-based representation. Each field boundary was placed into one text file, so the dataset contains 222 files with different field boundaries in GeoJSON (file name pattern ``\texttt{<PREFIX>.expected.txt}'' and 222 files containing the corresponding versions of the same field boundaries in John Deere's representation (``\texttt{<PREFIX>.input.txt}''). In addition, there is one \texttt{target.txt} file with an example of the expected output in GeoJSON. Furthermore, we created four versions of the dataset, with the purpose of increasing the complexity of the task:

\begin{itemize}
    \item \textbf{dataset-v1:} The GeoJSON files in this dataset (i.e., the expected output files) contain only the field boundaries, with no additional properties. That means that the conversion is expected to ignore additional properties that are present in the input files.
    \item \textbf{dataset-v2:} The GeoJSON files in this dataset contain, in addition to the field boundaries, the property \textit{id}, which is expected to be identified and brought from the input files. Note that the property id has the same representation in the input and in the expected output.
    \item \textbf{dataset-v3:} The GeoJSON files in this dataset contain, in addition to the field boundaries and id, the property \textit{area\_ha} (for area in hectares), which is expected to be identified and brought from the input files. Note that both the input and the expected output files have this properties, however they represent them differently.
    \item \textbf{dataset-v4} The GeoJSON files in this dataset contain, in addition to the field boundaries and id, the property \textit{area\_acres} (for area in acres), which is expected to be identified and converted from the input files. Here, the models also have to convert the area unit from hectares to acres.
    
\end{itemize}

Listing~\ref{lis:john-deere} shows an excerpt of the John Deere-based representation (input), whereas Listing~\ref{lis:geojson} does the same for the GeoJSON (expected output). These excerpts come from dataset-v4. Note, for example, the difference between the representation of field area and field boundaries in the input (lines 9-11 and 19-35) and in the expected output (lines 8 and 10-21).

\begin{lstlisting}[language=json, caption=Excerpt of an example of the input data, label={lis:john-deere}]
  "values": [
    {
      "@type": "Boundary",
      "id": "e2a217d3-d261-4f1b-9a7e-a719002ed933",
      "name": "Unique_Boundary_name",
      "sourceType": "HandDrawn",
      "createdTime": "2018-07-01T21:00:11Z",
      "modifiedTime": "2018-11-16T15:43:27.496Z",
      "area": {
        "@type": "MeasurementAsDouble",
        "valueAsDouble": 0.0921547167479482,
        "unit": "ha"
      },
      "workableArea": {
        "@type": "MeasurementAsDouble",
        "valueAsDouble": 0.0921547167479482,
        "unit": "ha"
      },
      "multipolygons": [
        {
          "@type": "Polygon",
          "rings": [
            {
              "@type": "Ring",
              "points": [
                {
                  "@type": "Point",
                  "lat": 52.330802,
                  "lon": 10.16014
                },
                {
                  "@type": "Point",
                  "lat": 52.330026,
                  "lon": 10.155896
                },
\end{lstlisting}

\begin{lstlisting}[language=json, caption=Excerpt of an example of the expected output data, label={lis:geojson}]
  {
  "type": "FeatureCollection",
  "features": [
    {
      "type": "Feature",
      "properties": {
        "id": "e2a217d3-d261-4f1b-9a7e-a719002ed933",
        "area_acres": 0.24787540406938302
      },
      "geometry": {
        "type": "Polygon",
        "coordinates": [
          [
            [
              10.16014,
              52.330802
            ],
            [
              10.155896,
              52.330026
            ],
\end{lstlisting}

\subsection{Instrumentation}\label{sec:instrumentation}

We created an evaluation program in Java that uses Ollama\footnote{\url{https://ollama.com}} as a platform to run the LLMs. Note that we did not use Ollama's feature to produce structured output in our evaluation for two reasons. First, as we wanted to evaluate the LLMs, adding an Ollama-specific feature that manipulates the structure of the models' output could act as a confounding  factor. Second, a solution that used structured outputs would be limited to produce output in JSON.

Ollama was installed locally on a hardware with three NVIDIA Tesla V100 SXM3 32 GB GPUs. The evaluation program receives as input a set of commands that allow using the models to process the datasets and post-processing the results. Input parameters include the location of the dataset, the url of the Ollama server that will run the model, the name of the model, the location of the results, and which strategy should be used (DIRECT or CODEGEN). Each strategy has a template prompt. These prompts were iteratively refined through trial-and-error\footnote{For this we used small models (up to 8B parameters): qwen2.5-coder:7b, codellama:7b, codeqwen:7b, deepseek-coder:6.7b, qwen2.5-coder:7b, codegemma:7b, and deepseek-r1:8b. Some of them entered the evaluation either as they are or in larger versions.} with small portions of the dataset until they reached their final forms. For the CODEGEN strategy, we prompt the models to generate code in Python 3 using only the Python Standard Library\footnote{\url{https://docs.python.org/3/library/index.html}}. Each prompt is sent to the model individually, i.e., no context is shared between the calls. Therefore, all measurements were based on zero-shot sampling. The post-processing commands include exporting the results to csv, which indicate the result of each attempt to convert data, including additional information in case of failure. %The program also generates shell scripts for program-independent reproduction of the experiment. 
To analyze the collected data, we used R\footnote{https://www.r-project.org/}, a free software for statistical computing.

\subsection{Model selection}\label{sec:model-selection}

We used the EvalPlus Leaderboard\footnote{\url{https://evalplus.github.io/leaderboard.html}} to select the models to be used in the experimentation, as they are an open and popular ranking among code-generation LLMs and use Python-related coding challenges as benchmarks \cite{liu2023your}. As the reproducibility of experiments with proprietary models can be limited \cite{wagner2025towards}, we selected only open-source models. Due to software and hardware constraints of our experimentation environment, we also limited the selection to models up to 70 billion parameters and that could be executed on the Ollama platform (see Section~\ref{sec:instrumentation}). Finally, as the EvalPlus Leaderboard ranks more than 120 models, we limited the scope of our evaluation to the models that scored above 0.7 on pass@1.

\subsection{Parameters, hypotheses, and variables}

\subsubsection{Parameters}

For checking the consistency across multiple runs, we limited the number of runs to 3. The temperature of all models has been set to 0.9 in all executions. As we selected models according to the previously mentioned ranking, we have manipulated neither their number of parameters nor their quantization (the values are reported on Section~\ref{sec:experiment-execution}).

\subsubsection{Hypotheses}

The following hypotheses have been defined and tested for both strategies:

\begin{enumerate}

    \item [H1] \textbf{Model comparison.} The average effectiveness of each model are compared to one another to check whether there is any significant difference among them. 
    \begin{itemize}
        \item $H1_0$: There is no difference in the average effectiveness of the selected LLMs.
        \item $H1_1$: There is a difference in the average effectiveness of the selected LLMs.
        
    \end{itemize}

    \item [H2] \textbf{Consistency check.} As LLMs are a non-deterministic technology, their results may vary across multiple executions. We want to check for each selected model whether eventual differences (if any) are significant. 
    \begin{itemize}
        \item $H2_0$: There is no difference in the effectiveness of LLMs across multiple runs.
        \item $H2_1$: There is a difference in the effectiveness of LLMs across multiple runs.
        
    \end{itemize}

    \item [H3] \textbf{Dataset comparison.} The average effectiveness of each model when processing different versions of the dataset is compared to check whether there is any significant difference among them. 
    \begin{itemize}
        \item $H3_0$: There is no difference in the average effectiveness of the selected LLMs using different dataset versions.
        \item $H3_1$: There is a difference in the average effectiveness of the selected LLMs using different dataset versions.
        
    \end{itemize}
\end{enumerate}

\subsection{Data collection and analysis procedure}

\begin{figure*}[]
  \centering
  \includegraphics[width=\linewidth]{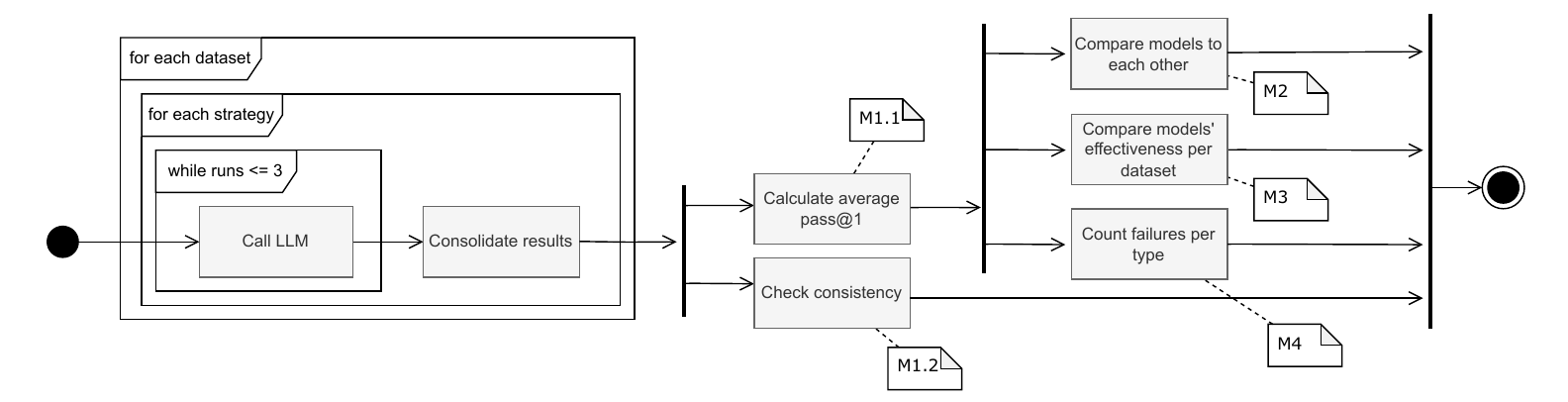}
  \caption{Experimental procedure.}
  \Description{Description of the experimental procedure figure.}
  \label{fig:procedure}
\end{figure*}

For each dataset and for each model, we performed three runs using both strategies. After each run, we calculated the corresponding pass@1; after all executions, we calculated the average pass@1 of each model across the three runs (M1.1). Then, for each dataset and strategy, we compared the average pass@1 of the models to identify which model performed better, if any(M1.2). Next, we checked whether the results of each model for each dataset were consistent across the runs (M2). Subsequently, we compared the models' effectiveness across the different dataset versions (M3). Finally, we counted the number of failures in the process, grouped by failure type (M4). Figure~\ref{fig:procedure} illustrates the procedure. To analyze the collected data, we used quantitative methods such as descriptive statistics and hypothesis testing (for comparisons). For the statistical analysis, we have set the significance level to $\alpha = 0.05$ and $\beta = 0.2$. Cohen's $h$ has been used to calculate the effect sizes (in R, \texttt{ES.h}). For measuring power, we applied the function \texttt{pwr.2p.test} from the R package \texttt{pwr}.

\section{Results}\label{sec:results}

\subsection{Experiment execution}\label{sec:experiment-execution}

Following the criteria described in Section~\ref{sec:model-selection}, thirteen models have been selected on June 6, 2025, to participate in the experiment. Table~\ref{tab:models} list the models with their corresponding checksums, model sizes, and quantization schemas.% A series of shell scripts have been prepared to make successive calls to the evaluation program to execute all tests. 

\begin{table}[b]
\caption{Selected models ranked by EvalPlus scores.}
\label{tab:models}
\begin{tabular}{lrll}
\hline
\textbf{Model tag}              & \multicolumn{1}{c}{\textbf{Size}} & \textbf{Quant.} & \textbf{Checksum} \\ \hline
qwen2.5-coder:32b               & 32B                               & Q4\_K\_M        & b92d6a0bd47e      \\
deepseek-coder-v2:16b           & 16B                               & Q4\_0           & 63fb193b3a9b      \\
codeqwen:7b                     & 7B                                & Q4\_0           & df352abf55b1      \\
opencoder:8b                    & 8B                                & Q4\_K\_M        & cd882db52297      \\
deepseek-coder:33b              & 33B                               & Q4\_0           & acec7c0b0fd9      \\
codestral:22b                   & 22B                               & Q4\_0           & 0898a8b286d5      \\
wojtek/opencodeinterpreter:33b  & 33B                               & Q5\_K\_M        & e293532d0e21      \\
wizardcoder:33b                 & 33B                               & Q4\_0           & 5b944d3546ee      \\
llama3:70b                      & 70B                               & Q4\_0           & 786f3184aec0      \\
mixtral:8x22b                   & 141B                              & Q4\_0           & e8479ee1cb51      \\
wojtek/opencodeinterpreter:6.7b & 6.7B                              & Q8\_0           & 50f75db69081      \\
deepseek-coder:6.7b             & 6.7B                              & Q4\_0           & ce298d984115      \\
limsag/starchat2:15b-v0.1-f16   & 15B                               & F16             & 5a36458e440e      \\ \hline
\end{tabular}
\end{table}

\subsubsection{Deviation 1:} During executions, we noticed that two models were taking too long to generate results: mixtral:8x22b and opencoder:8b. In the first case, we realized that the model has been selected accidentally, as it has 141B. For this reason, we stopped the evaluation of this model and omitted the partial results from this report. The other case, opencoder:8b, presented bad performance only using the strategy DIRECT (nearly 20 minutes, whereas most models return within a couple of minutes or even a few seconds, in some cases). The preliminary 58 results, obtained from dataset v1, showed that opencoder:8b was failing in all attempts. In 56 of them, the error was ``LLM exception: (`Completion stopped abnormally: length')'' (see different failure causes in Section~\ref{sec:failurecauses}). For this reason, we stopped the evaluation of opencoder:8b using the strategy DIRECT and omitted the partial results from this report (the results obtained using CODEGEN have been kept).

\subsubsection{Deviation 2:} After the executions, the logs indicate that some failures happened due to network connections. These results have been deleted, and the run has been executed again.

\subsubsection{Deviation 3:} After the executions on the dataset v4, which includes a unit conversion step, we noticed that all models failed. Checking the logs, we noticed that in numerous instances the difference between the expected result and the generated result was exactly in the property that has a converted unit. The results were similar; however, there were differences in the late decimal places of the numbers. As we reviewed the dataset generation procedure, we identified a problem when we created the dataset v4: we performed multiple float point conversions among the different units in the dataset preparation process, which led to losing the precision of the numbers. For this reason, we regenerated the dataset v4 performing only one unit conversion (from the input unit to the expected unit) and re-run the experiment on this dataset.

\subsection{Analysis}

In total, we performed nearly 64.000 LLM calls (2 strategies $\times$ 4 dataset versions $\times$ 222 entries $\times$ 12 models $\times$ 3 runs). \textbf{Table~\ref{tab:results} consolidates the results of processing all datasets using the selected models and both strategies}. Due to space constraints, we did not reproduce the results of all statistical tests in the paper\footnote{All statistical tests and results can be found in the experimentation package.}.

\subsubsection{On the models' effectiveness (Q1)}

\textbf{Using the strategy DIRECT}, \qwen~ outperformed by far all other models. It scored an average pass@1 > 0.99 in datasets v1, v2, and v3, whereas most of other models stayed below 0.2. When processing v4, \qwen~(and all other models) failed. The difference between the effectiveness of \qwen~ and the second-best models was significant in all dataset versions but v4. That means, using v1, v2, and v3, $H1_0$ can be rejected when comparing \qwen~ with other models, and $H1_1$ can be accepted. All tests comparing \qwen~ with the second-best models showed enough power and high effect sizes. One interesting aspect observed in the results of \qwen~ in dataset v1, where there were only two failures, was that these failures occurred due to a rounding problem in a float number representing the field area.
%\subsubsection{CODEGEN}
\textbf{As for the strategy CODEGEN}, the most effective model across all datasets was, once again, qwen2.5-coder:32b. The average pass@1 were 0.8948949 (v1), 0.9129129 (v2),  0.9309309 (v3), and 0.7522522 (v4). llama3:70b's shared the first place with qwen2.5-coder:32b in datasets v1 and v2 ($H1_0$ cannot be rejected, p-value=0.786 and 0.2613, respectively). In datasets v3, qwen2.5-coder:32b was found more effective, although with a small effect size and low power in comparison with \llama~ ($H1_0$ can be rejected, p-value=0.03185; $h=0.12$, power=0.615), whereas in v4, the effect size was large, as llama3:70b scored impressively low (average pass@1 = 0.0120120; $H2_2$ can be rejected, p-value < 2.2e-16; $H2_1$ can be accepted, $h=1.87$, power=1). Most of the other models performed considerably worse, with many of them not even reaching 0.5 in any evaluation.

\subsubsection{On the models' consistency (Q2)}

Most models were consistent regarding the results they produced in both strategies, as can be seen in Table~\ref{tab:results} (consistent results across multiple runs are written in bold font), despite of the non-deterministic nature of the technology and that we have set the temperature to 0.9 in all cases.

Consistency can also be regarded as another light advantage presented by qwen2.5-coder:32b in comparison with \llama~ using the strategy CODEGEN, as \qwen~ was consistent across the three runs: in all cases, no significant difference was found among the results of each run, whereas llama3:70b was inconsistent when processing v1 (low effect size and no enough power, though: in the comparison between runs 2 and 3, $H2_0$ can be rejected, p-value=0.01358; however, $H2_1$ cannot be accepted: effect size $h=0$, power=0.05) and v2 (low effect size: for runs 1 and 3, where $H2_0$ can be rejected, p-value=0.03754; $H2_1$ can be accepted, effect size $h=0.21$, power=0.97).

% Please add the following required packages to your document preamble:
% \usepackage{booktabs}
% \usepackage{multirow}
\begin{table*}[]
\caption{Effectiveness of models (DIRECT/CODEGEN) on datasets (v1-v4). Gray cells mark the best model per strategy/dataset. Bold pass@1 indicates consistent results across runs.}
\label{tab:results}
\resizebox{\linewidth}{!}{%
\begin{tabular}{ll|rrrr|rrrr}
\hline
                                   &                                      & \multicolumn{4}{c|}{\textbf{pass@1 DIRECT}}                                                                                                               & \multicolumn{4}{c}{\textbf{pass@1 CODEGEN}}                                                                                                               \\ \cline{3-10} 
\multirow{-2}{*}{\textbf{Dataset}} & \multirow{-2}{*}{\textbf{Model tag}} & \multicolumn{1}{c}{\textit{Run 1}} & \multicolumn{1}{c}{\textit{Run 2}} & \multicolumn{1}{c}{\textit{Run 3}} & \multicolumn{1}{c|}{\textit{Average}}      & \multicolumn{1}{c}{\textit{Run 1}} & \multicolumn{1}{c}{\textit{Run 2}} & \multicolumn{1}{c}{\textit{Run 3}} & \multicolumn{1}{c}{\textit{Average}}       \\ \hline
v1                                 & qwen2.5-coder:32b                    & \textbf{1}                         & \textbf{0.9954955}                 & \textbf{0.9954955}                 & \cellcolor[HTML]{C0C0C0}\textbf{0.9969969} & \textbf{0.8918919}                 & \textbf{0.9144144}                 & \textbf{0.8783784}                 & \cellcolor[HTML]{C0C0C0}\textbf{0.8948949} \\
                                   & deepseek-coder-v2:16b                & \textbf{0.0675675}                 & \textbf{0.0855855}                 & \textbf{0.0855855}                 & \textbf{0.0795795}                         & \textbf{0.7162162}                 & \textbf{0.7612613}                 & \textbf{0.7747748}                 & \textbf{0.7507508}                         \\
                                   & codeqwen:7b                          & \textbf{0.0225225}                 & \textbf{0.0360360}                 & \textbf{0.0270270}                 & \textbf{0.0285285}                         & \textbf{0.3783784}                 & \textbf{0.3513514}                 & \textbf{0.3378378}                 & \textbf{0.3558559}                         \\
                                   & opencoder:8b                         & N/A                                & N/A                                & N/A                                & N/A                                        & \textbf{0.1801802}                 & \textbf{0.2162162}                 & \textbf{0.1711712}                 & \textbf{0.1891892}                         \\
                                   & deepseek-coder:33b                   & \textbf{0}                         & \textbf{0}                         & \textbf{0.0090090}                 & \textbf{0.0030030}                         & \textbf{0}                         & \textbf{0}                         & \textbf{0}                         & \textbf{0}                                 \\
                                   & codestral:22b                        & \textbf{0.0045045}                 & \textbf{0}                         & \textbf{0.0045045}                 & \textbf{0.0030030}                         & \textbf{0.6666666}                 & \textbf{0.6081081}                 & \textbf{0.5900900}                 & \textbf{0.6216216}                         \\
                                   & wojtek/opencodeinterpreter:33b       & \textbf{0}                         & \textbf{0}                         & \textbf{0}                         & \textbf{0}                                 & \textbf{0}                         & \textbf{0}                         & \textbf{0}                         & \textbf{0}                                 \\
                                   & wizardcoder:33b                      & \textbf{0.1801801}                 & \textbf{0.1306306}                 & \textbf{0.1441441}                 & \textbf{0.1516516}                         & \textbf{0.01801802}                & \textbf{0.01801802}                & \textbf{0.03153153}                & \textbf{0.02252252}                        \\
                                   & llama3:70b                           & 0.1607142                          & 0                                  & 0                                  & 0.0535714                                  & 0.9189189                          & 0.9324324                          & 0.8558558                          & \cellcolor[HTML]{C0C0C0}0.9024024          \\
                                   & wojtek/opencodeinterpreter:6.7b      & \textbf{0.0270270}                 & \textbf{0.0270270}                 & \textbf{0.0180180}                 & \textbf{0.0240240}                         & \textbf{0.2657658}                 & \textbf{0.3108108}                 & \textbf{0.2972973}                 & \textbf{0.2912913}                         \\
                                   & deepseek-coder:6.7b                  & \textbf{0.0630630}                 & \textbf{0.0450450}                 & \textbf{0.0675675}                 & \textbf{0.0585585}                         & \textbf{0.5495495}                 & \textbf{0.6171171}                 & \textbf{0.5675676}                 & \textbf{0.5780781}                         \\
                                   & limsag/starchat2:15b-v0.1-f16        & \textbf{0}                         & \textbf{0}                         & \textbf{0}                         & \textbf{0}                                 & \textbf{0.2792793}                 & \textbf{0.3243243}                 & \textbf{0.3198198}                 & \textbf{0.3078078}                         \\ \hline
v2                                 & qwen2.5-coder:32b                    & \textbf{0.9954954}                 & \textbf{0.9954954}                 & \textbf{0.9954954}                 & \cellcolor[HTML]{C0C0C0}\textbf{0.9954954} & \textbf{0.9234234}                 & \textbf{0.9324324}                 & \textbf{0.8828829}                 & \cellcolor[HTML]{C0C0C0}\textbf{0.9129129} \\
                                   & deepseek-coder-v2:16b                & \textbf{0.6486486}                 & \textbf{0.5765765}                 & \textbf{0.6261261}                 & \textbf{0.6171171}                         & \textbf{0.5720720}                 & \textbf{0.6486486}                 & \textbf{0.6081081}                 & \textbf{0.6096096}                         \\
                                   & codeqwen:7b                          & \textbf{0.0135135}                 & \textbf{0.0090090}                 & \textbf{0}                         & \textbf{0.0075075}                         & \textbf{0.0720720}                 & \textbf{0.06306306}                & \textbf{0.06756757}                & \textbf{0.06756757}                        \\
                                   & opencoder:8b                         & N/A                                & N/A                                & N/A                                & N/A                                        & \textbf{0.1486486}                 & \textbf{0.1441441}                 & \textbf{0.1396396}                 & \textbf{0.1441441}                         \\
                                   & deepseek-coder:33b                   & \textbf{0.0090090}                 & \textbf{0.0045045}                 & \textbf{0.0045045}                 & \textbf{0.0060060}                         & \textbf{0}                         & \textbf{0}                         & \textbf{0}                         & \textbf{0}                                 \\
                                   & codestral:22b                        & 0.5765765                          & 0.5135135                          & 0.4729729                          & 0.5210210                                  & \textbf{0.6441441}                 & \textbf{0.5810811}                 & \textbf{0.6441441}                 & \textbf{0.6231231}                         \\
                                   & wojtek/opencodeinterpreter:33b       & \textbf{0}                         & \textbf{0}                         & \textbf{0}                         & \textbf{0}                                 & \textbf{0}                         & \textbf{0}                         & \textbf{0}                         & \textbf{0}                                 \\
                                   & wizardcoder:33b                      & \textbf{0.1666666}                 & \textbf{0.1756756}                 & \textbf{0.1576576}                 & \textbf{0.1666666}                         & \textbf{0}                         & \textbf{0}                         & \textbf{0}                         & \textbf{0}                                 \\
                                   & llama3:70b                           & 0.0945945                          & 0.1801801                          & 0.1306306                          & 0.1351351                                  & 0.9594595                          & 0.9279279                          & 0.9054054                          & \cellcolor[HTML]{C0C0C0}0.9309309          \\
                                   & wojtek/opencodeinterpreter:6.7b      & \textbf{0.0045045}                 & \textbf{0}                         & \textbf{0}                         & \textbf{0.0015015}                         & \textbf{0.2252252}                 & \textbf{0.2207207}                 & \textbf{0.1981982}                 & \textbf{0.2147147}                         \\
                                   & deepseek-coder:6.7b                  & \textbf{0.0090090}                 & \textbf{0}                         & \textbf{0}                         & \textbf{0.0030030}                         & \textbf{0.4864865}                 & \textbf{0.545045}                  & \textbf{0.509009}                  & \textbf{0.5135135}                         \\
                                   & limsag/starchat2:15b-v0.1-f16        & \textbf{0}                         & \textbf{0}                         & \textbf{0}                         & \textbf{0}                                 & \textbf{0.0495495}                 & \textbf{0.0720721}                 & \textbf{0.0405405}                 & \textbf{0.0540541}                         \\ \hline
v3                                 & qwen2.5-coder:32b                    & \textbf{1}                         & \textbf{1}                         & \textbf{1}                         & \cellcolor[HTML]{C0C0C0}\textbf{1}         & \textbf{0.9414414}                 & \textbf{0.9324324}                 & \textbf{0.9189189}                 & \cellcolor[HTML]{C0C0C0}\textbf{0.9309309} \\
                                   & deepseek-coder-v2:16b                & \textbf{0.8513513}                 & \textbf{0.8423423}                 & \textbf{0.8378378}                 & \textbf{0.8438438}                         & 0.509009                           & 0.5585586                          & 0.6216216                          & 0.5630631                                  \\
                                   & codeqwen:7b                          & \textbf{0.1081081}                 & \textbf{0.1171171}                 & \textbf{0.0855855}                 & \textbf{0.1036036}                         & \textbf{0.0855856}                 & \textbf{0.0720721}                 & \textbf{0.1126126}                 & \textbf{0.0900901}                         \\
                                   & opencoder:8b                         & N/A                                & N/A                                & N/A                                & N/A                                        & \textbf{0.2162162}                 & \textbf{0.1441441}                 & \textbf{0.1801802}                 & \textbf{0.1801802}                         \\
                                   & deepseek-coder:33b                   & \textbf{0}                         & \textbf{0.0045045}                 & \textbf{0}                         & \textbf{0.0015015}                         & \textbf{0}                         & \textbf{0}                         & \textbf{0}                         & \textbf{0}                                 \\
                                   & codestral:22b                        & \textbf{0.7927927}                 & \textbf{0.7657657}                 & \textbf{0.7837837}                 & \textbf{0.7807807}                         & \textbf{0.6846847}                 & \textbf{0.6306306}                 & \textbf{0.6621622}                 & \textbf{0.6591592}                         \\
                                   & wojtek/opencodeinterpreter:33b       & \textbf{0}                         & \textbf{0}                         & \textbf{0}                         & \textbf{0}                                 & \textbf{0}                         & \textbf{0}                         & \textbf{0}                         & \textbf{0}                                 \\
                                   & wizardcoder:33b                      & \textbf{0.1081081}                 & \textbf{0.1396396}                 & \textbf{0.0900900}                 & \textbf{0.1126126}                         & \textbf{0}                         & \textbf{0}                         & \textbf{0}                         & \textbf{0}                                 \\
                                   & llama3:70b                           & \textbf{0.8108108}                 & \textbf{0.8558558}                 & \textbf{0.8738738}                 & \textbf{0.8468468}                         & \textbf{0.8918918}                 & \textbf{0.8963963}                 & \textbf{0.9009009}                 & \textbf{0.8963963}                         \\
                                   & wojtek/opencodeinterpreter:6.7b      & \textbf{0}                         & \textbf{0}                         & \textbf{0}                         & \textbf{0}                                 & \textbf{0.2522522}                 & \textbf{0.2207207}                 & \textbf{0.2072072}                 & \textbf{0.2267267}                         \\
                                   & deepseek-coder:6.7b                  & \textbf{0.0225225}                 & \textbf{0.0315315}                 & \textbf{0.0045045}                 & \textbf{0.0195195}                         & \textbf{0.4594594}                 & \textbf{0.5045045}                 & \textbf{0.4819819}                 & \textbf{0.4819819}                         \\
                                   & limsag/starchat2:15b-v0.1-f16        & \textbf{0}                         & \textbf{0}                         & \textbf{0}                         & \textbf{0}                                 & \textbf{0.0675675}                 & \textbf{0.0450450}                 & \textbf{0.0450450}                 & \textbf{0.0525525}                         \\ \hline
v4                                 & qwen2.5-coder:32b                    & \textbf{0}                         & \textbf{0}                         & \textbf{0}                         & \textbf{0}                                 & \textbf{0.7162162}                 & \textbf{0.7792792}                 & \textbf{0.7612612}                 & \cellcolor[HTML]{C0C0C0}\textbf{0.7522522} \\
                                   & deepseek-coder-v2:16b                & \textbf{0}                         & \textbf{0}                         & \textbf{0}                         & \textbf{0}                                 & \textbf{0.0855855}                 & \textbf{0.0810810}                 & \textbf{0.0945945}                 & \textbf{0.0870870}                         \\
                                   & codeqwen:7b                          & \textbf{0}                         & \textbf{0}                         & \textbf{0}                         & \textbf{0}                                 & \textbf{0}                         & \textbf{0}                         & \textbf{0}                         & \textbf{0}                                 \\
                                   & opencoder:8b                         & N/A                                & N/A                                & N/A                                & N/A                                        & \textbf{0.0045045}                 & \textbf{0.0225225}                 & \textbf{0.0135135}                 & \textbf{0.0135135}                         \\
                                   & deepseek-coder:33b                   & \textbf{0}                         & \textbf{0}                         & \textbf{0}                         & \textbf{0}                                 & \textbf{0}                         & \textbf{0}                         & \textbf{0}                         & \textbf{0}                                 \\
                                   & codestral:22b                        & \textbf{0}                         & \textbf{0}                         & \textbf{0}                         & \textbf{0}                                 & \textbf{0.2792792}                 & \textbf{0.2927927}                 & \textbf{0.2927927}                 & \textbf{0.2882882}                         \\
                                   & wojtek/opencodeinterpreter:33b       & \textbf{0}                         & \textbf{0}                         & \textbf{0}                         & \textbf{0}                                 & \textbf{0}                         & \textbf{0}                         & \textbf{0}                         & \textbf{0}                                 \\
                                   & wizardcoder:33b                      & \textbf{0}                         & \textbf{0}                         & \textbf{0}                         & \textbf{0}                                 & \textbf{0}                         & \textbf{0}                         & \textbf{0}                         & \textbf{0}                                 \\
                                   & llama3:70b                           & \textbf{0}                         & \textbf{0}                         & \textbf{0}                         & \textbf{0}                                 & \textbf{0}                         & \textbf{0.0225225}                 & \textbf{0.0135135}                 & \textbf{0.0120120}                         \\
                                   & wojtek/opencodeinterpreter:6.7b      & \textbf{0}                         & \textbf{0}                         & \textbf{0}                         & \textbf{0}                                 & \textbf{0.0270270}                 & \textbf{0.0090090}                 & \textbf{0.0180180}                 & \textbf{0.0180180}                         \\
                                   & deepseek-coder:6.7b                  & \textbf{0}                         & \textbf{0}                         & \textbf{0}                         & \textbf{0}                                 & \textbf{0.0315315}                 & \textbf{0.0180180}                 & \textbf{0.0045045}                 & \textbf{0.0180180}                         \\
                                   & limsag/starchat2:15b-v0.1-f16        & \textbf{0}                         & \textbf{0}                         & \textbf{0}                         & \textbf{0}                                 & \textbf{0.0045045}                 & \textbf{0.0045045}                 & \textbf{0}                         & \textbf{0.0030030}                         \\ \hline
\end{tabular}%
}
\end{table*}

\subsubsection{On the influence of the dataset versions (Q3)}
\begin{figure}[]
  \centering
  \includegraphics[width=\linewidth]{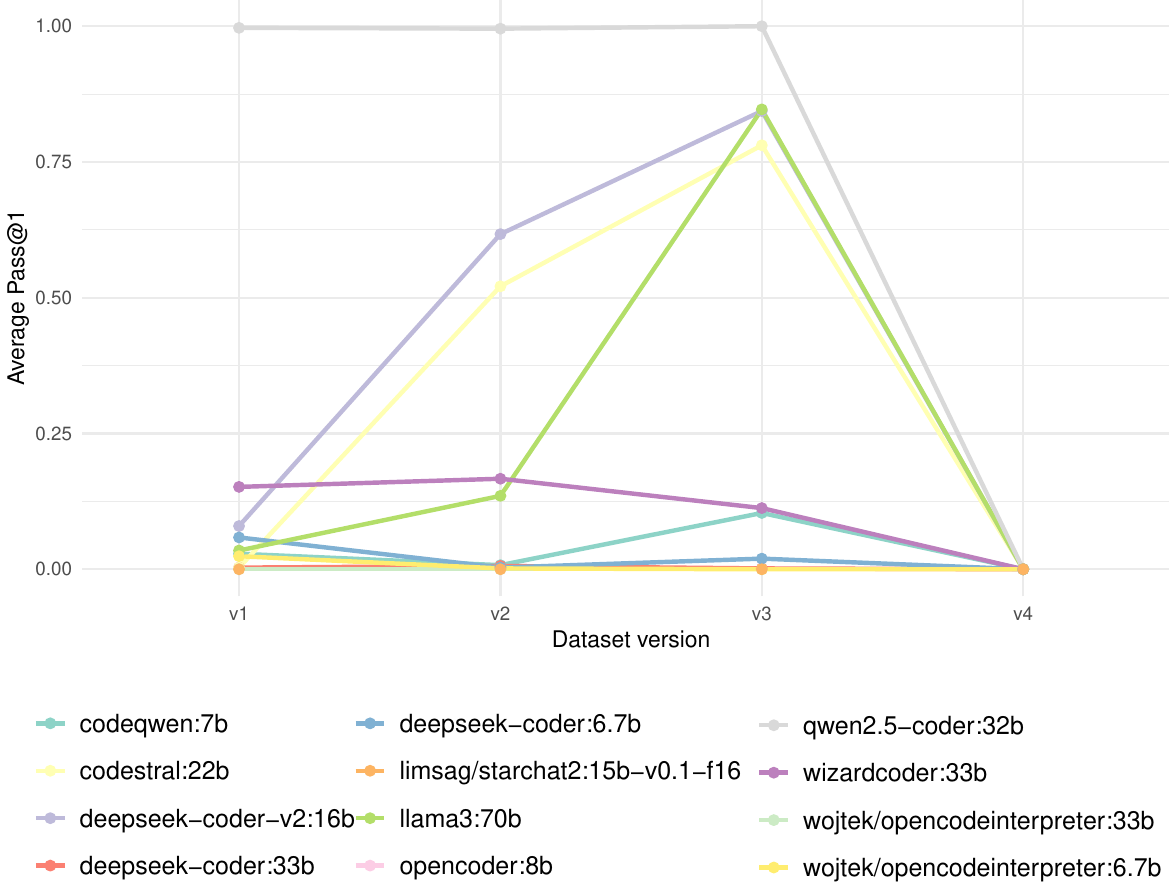}
  \caption{Comparison of average pass@1 of the models using DIRECT across the four dataset versions.}
  \Description{Comparison of average pass@1 of the models using DIRECT across the four dataset versions.}
  \label{fig:direct-dataset-comparison}
\end{figure}

\begin{figure}[]
  \centering
  \includegraphics[width=\linewidth]{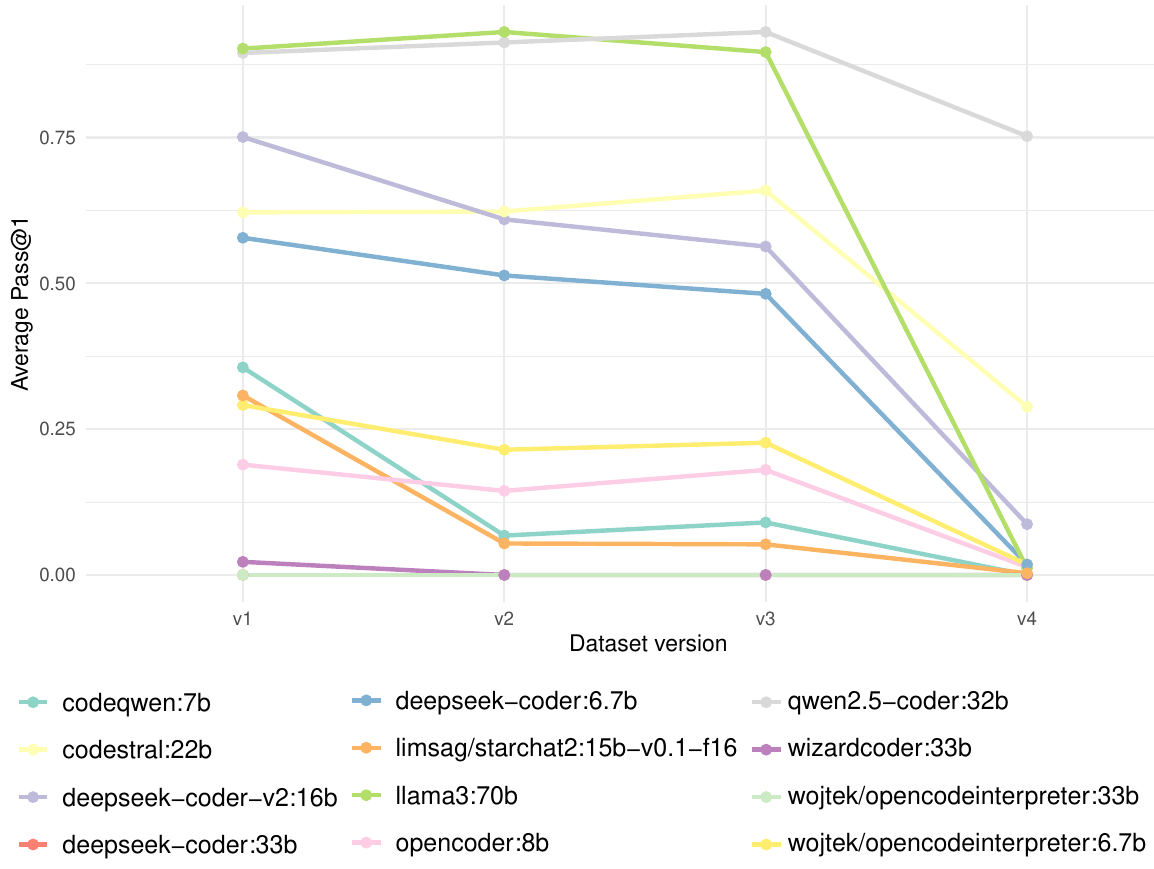}
  \caption{Comparison of average pass@1 of the models using CODEGEN across the four dataset versions.}
  \Description{Comparison of average pass@1 of the models using CODEGEN across the four dataset versions.}
  \label{fig:codegen-dataset-comparison}
\end{figure}

\textbf{Using DIRECT }the dataset version had an unexpected impact on certain models: as the dataset version became more complex, many models performed significantly better. For example, when processing v1, deepseek-coder-v2:16b scored a consistent average pass@1 = 0.07; when processing v2, 0.61; finally, for v3, 0.84. In all cases $H3_0$ can be rejected, and the comparisons had a medium or large effect size and enough power. Upon review, many v1 failures likely occurred because the model added extra key-value pairs from the input to the ``properties'' attribute, which were not part of the target representation. The same phenomenon was observed with other models (see codestral:22b, llama3:70b). Conversely, when processing the dataset v4, none of the models were able to convert the data as expected, i.e., all executions failed. As mentioned in Section~\ref{sec:dataset-prep}, this dataset required, for a certain property, not only a change in the data structure but also a unit conversion. Figure~\ref{fig:direct-dataset-comparison} illustrate the results. 
\textbf{Using CODEGEN}, some models vary in their effectiveness among the dataset versions v1, v2, and v3 (e.g., deepseek-coder-v2:16b was more effective processing dataset v1 than v2, i.e., $H3_0$ has been rejected), however the major different was observed the results produced by the models using the dataset v4. As can be seen in Figure~\ref{fig:codegen-dataset-comparison}, all models -- noticeably the best performing models -- were less effective when using v4, including \qwen.

Table~\ref{tab:dataset-comparison} shows the comparison across dataset versions processed by \qwen~ using both strategies. As can be seen, while adding new properties to the dataset without changing the units did not bring issues to the model effectiveness, the introduction of unit convertion had a measurable impact.

\begin{table}[]
\caption{Effectiveness of \qwen~across datasets. Gray cells show when $H3_0$ is rejected with sufficient power.}
\label{tab:dataset-comparison}
\begin{tabular}{@{}llrrr@{}}
\toprule
Strategy & \multicolumn{1}{c}{Comparison}   & \multicolumn{1}{c}{\begin{tabular}[c]{@{}c@{}}Z-test\\ (p-value)\end{tabular}} & \multicolumn{1}{c}{\begin{tabular}[c]{@{}c@{}}Effect size\\ ($h$)\end{tabular}} & \multicolumn{1}{c}{\begin{tabular}[c]{@{}c@{}}Power\\ ($1-\beta$)\end{tabular}} \\ \midrule

DIRECT   & v1 vs v2                         & 1                                                                              & -                                                                             & -                                                                            \\
         & v1 vs v3                         & 0.4792                                                                         & -                                                                             & -                                                                            \\
         & \cellcolor[HTML]{C0C0C0}v1 vs v4 & \cellcolor[HTML]{C0C0C0}< 2.2e-16                                              & \cellcolor[HTML]{C0C0C0}3.03                                                  & \cellcolor[HTML]{C0C0C0}1                                                    \\
         & v2 vs v3                         & 0.2477                                                                         & -                                                                             & -                                                                            \\
         & \cellcolor[HTML]{C0C0C0}v2 vs v4 & \cellcolor[HTML]{C0C0C0}< 2.2e-16                                              & \cellcolor[HTML]{C0C0C0}3.00                                                  & \cellcolor[HTML]{C0C0C0}1                                                    \\
         & \cellcolor[HTML]{C0C0C0}v3 vs v4 & \cellcolor[HTML]{C0C0C0}< 2.2e-16                                              & \cellcolor[HTML]{C0C0C0}3.14                                                  & \cellcolor[HTML]{C0C0C0}1                                                    \\ \midrule
CODEGEN  & v1 vs v2                         & 0.3065                                                                         & -                                                                             & -                                                                            \\
         & v1 vs v3                         & 0.02542                                                                        & -0.12                                                                         & 0.64                                                                         \\
         & \cellcolor[HTML]{C0C0C0}v1 vs v4 & \cellcolor[HTML]{C0C0C0}1.41e-11                                               & \cellcolor[HTML]{C0C0C0}0.38                                                  & \cellcolor[HTML]{C0C0C0}0.99                                                 \\
         & v2 vs v3                         & 0.2613                                                                         & -                                                                             & -                                                                            \\
         & \cellcolor[HTML]{C0C0C0}v2 vs v4 & \cellcolor[HTML]{C0C0C0}7.293e-15                                              & \cellcolor[HTML]{C0C0C0}0.44                                                  & \cellcolor[HTML]{C0C0C0}0.99                                                 \\
         & \cellcolor[HTML]{C0C0C0}v3 vs v4 & \cellcolor[HTML]{C0C0C0}< 2.2e-16                                              & \cellcolor[HTML]{C0C0C0}0.51                                                  & \cellcolor[HTML]{C0C0C0}0.99                                                 \\ \bottomrule
\end{tabular}
\end{table}

\subsubsection{On failure causes (Q4)}\label{sec:failurecauses}

\textbf{In DIRECT}, almost 75\% all failures occurred due to either mismatches between the generated and the expected output or the generated JSON was syntactically incorrect. The next most frequent cause was runtime error (26.1\%) produced by Ollama with certain models (all except two being caused by the models deepseek-coder:33b, wizardcoder:33b, and wojtek/opencodeinterpreter:33b) \textit{when using our prompts} (we tested other simple prompts and the models produced responses without problems). In these cases, it is not clear whether the problem was rooted in the models, in Ollama, or a combination of both.
\textbf{In CODEGEN}, the most frequent failure cause was the mismatch between the JSON data produced by the LLM-generated code and the expected JSON data (37.2\%), followed by the runtime error (34.1\%) previously mentioned. \qwen, the most effective model, failed in 12.7\% of the calls -- nearly half of them (48.6\%) when processing the dataset v4. The majority of \qwen's failures were due to the generated JSON did not match the expected JSON (92\%), followed by code execution problems (0.05\%) and code compilation problems (0.02\%). Table~\ref{tab:fail-codegen} aggregates the occurrences for each identified failure cause in all LLM calls.

\begin{table}[]
\caption{Failure causes using DIRECT (N=24339).}
\label{tab:fail-direct}
\begin{tabular}{@{}lp{0.3\textwidth}rr@{}}
\toprule
\# & \textbf{Failure}                                         & \textbf{Count} & \textbf{\%}      \\ \midrule
1  & JSON data mismatches expected          & 9062          & 37.2             \\
%\rowcolor[HTML]{EFEFEF} 
2  & JSON syntax exception                                    & 8867           & 36.4             \\
3  & Runtime Exception when calling the LLM                   & 6353          & 26.1             \\
%\rowcolor[HTML]{EFEFEF} 
4  & LLM exception ("Completion stopped abnormally: length")  & 57             & 0.2            \\ \bottomrule
\end{tabular}
\end{table}

\begin{table}[]
\caption{Failure causes using CODEGEN (N=22472).}
\label{tab:fail-codegen}
\begin{tabular}{@{}lp{0.3\textwidth}rr@{}}
\toprule
\# & \textbf{Failure}                                         & \textbf{Count} & \textbf{\%}      \\ \midrule

1  & Runtime Exception when calling the LLM                   & 7950          & 35.4             \\
%\rowcolor[HTML]{EFEFEF} 
2  & JSON data mismatches expected         & 7838          & 34.9             \\
3  & Code execution exception & 4109           & 18.3             \\
%\rowcolor[HTML]{EFEFEF} 
4  & Code compilation exception                               & 2325           & 10.3              \\
5  & Data is empty                                            & 232            & 1.0            \\
%\rowcolor[HTML]{EFEFEF} 
6  & JSON syntax exception                                    & 17           & 0.1             \\

7  & HTTP timeout                                             & 1              & \textless 0.01 \\ \bottomrule
\end{tabular}
\end{table}

\section{Interpretation and discussion}\label{sec:interpretation}

LLM-assisted coding has become increasingly popular lately to support development time activities, but would LLMs be able to help systems interoperate at runtime? From a functional perspective, our results indicate that some LLMs \textit{can} do it. The model \qwen~, using the strategy DIRECT, in the best evaluation scenario, scored an average pass@1 = 1 (v3) , whereas using CODEGEN it scored pass@1 = 0.93 at its best (also v3). In contrast, our results, which were limited to a selection of open source LLMs of size up to 70B, also show this task is not trivial for most models, though. While all selected LLMs have scored above 0.7 on the EvalPlus Leaderboard, only a few achieve this threshold in our experiment. We see a need for better benchmarks for measuring the effectiveness of LLMs on data conversion tasks, as the results we found deviate largely from the one provided by the models on the reference benchmark. We expect our dataset work as an initial contribution to be reused in future evaluations. Ideally, there should be various datasets from different domains and varying levels of complexity -- always reflecting production use cases.

Selecting the best strategy to implement LLM-based interoperability involves a series of trade-offs. DIRECT has the advantages of being easier to implement, and although in some cases it can be even more effective than CODEGEN, it delivers an always non-deterministic solution. Every time a new data conversion is required, a model must be called (which also has computational and energy implications) -- let alone the fact that depending on the complexity of the data conversion, it may fail completely (see results of DIRECT using the dataset v4). On the other hand, CODEGEN requires more attention and efforts to implement, as code is generated, deployed, and executed on the fly. But from our perspective, the effort pays off, as the generated solution is deterministic and, when effective, can be reused with no additional need to call the LLMs again. We regard CODEGEN as the most promising strategy to develop the field of LLM-based interoperability further.

Making systems interoperate via LLM-based interfaces requires a certain effort, which includes (1) implementing the LLM-based component, (2) selecting (and potentially adjusting) base models, and (3) customizing prompts for the use case of interest. While the first step may be regarded as a one-time activity, the other might be needed on a case-by-case basis. Therefore, it is necessary to analyze empirically whether the investment to use LLM-based components for interoperability outweighs the efforts that a development team would need to do the same, which includes understanding foreign schemas, implementing adapters, testing, and deploying the solutions. We believe that the more heterogeneous and open the software-based ecosystem is, the greater the benefits provided by autonomous interoperability will be.

Finally, it is fair to ask in what use cases an effectiveness below 1 is acceptable. In many, including agriculture, it might not be. In our evaluation, the introduction of unit conversions (dataset v4) increased significantly the task difficulty. Rounding float numbers can also be an issue. From our perspective, beyond functional correctness, the next step is to make these solutions more reliable -- either by improving effectiveness or finding ways to identify and address failures.

\subsection{Threats to validity}

\subsubsection{Internal validity} We constrained the selection of models by our hardware and software infrastructure. A different selection would include some models that were better ranked at the EvalPlus Leaderboard, which could have provided better results in terms of effectiveness of the LLM-based strategies. Another point to the model selection is that we looked for models trained on code, but in a strict sense only the strategy CODEGEN generates code (strategy DIRECT generates structured data). Therefore, the results of strategy DIRECT could be different (better?) if other models had been selected. Moreover, other benchmarks than EvalPlus could have provided a better list of candidate models for experimentation. Concerning the type of data we used, previous knowledge about the data representations could influence the effectiveness of the models. By interacting with several models, we could not identify in them previous knowledge about the input data representation we used (based on John Deere's API). As for the target representation, GeoJSON is very well known. Regarding the consistency check, a larger number of runs could have yielded different results.

\subsubsection{External validity} While the non-deterministic nature of LLMs threatens the generalization of the results, we have mitigated it by checking the consistency of the LLMs across multiple runs. As for the suitability of LLMs to support the two presented strategies, even considering the most effective models, this evaluation was limited to a set of datasets in a specific use case. Results may vary when other use cases are considered.

\subsubsection{Construct validity} In our experiment, we used the same prompts to interact with all models. However, as LLMs are sensitive to the prompt they receive, different models may perform differently depending on the prompt they receive. Engineering individual prompts for each model could have led to different (better) results for some models. % This has not been investigated in the scope of this research. 

\subsubsection{Conclusion validity} In some comparisons that found significant differences between the models, the power was low, which brings a risk to these specific conclusions.

\section{Conclusion}\label{sec:conclusion}

In this paper, we contributed an empirical evaluation of two LLM-based strategies to make systems interoperate autonomously. The contribution includes four versions of manually curated datasets in an agricultural use case. The results indicate that some LLMs can make systems interoperate without previous knowledge of the incoming data, which can save development time efforts to implementing interoperability artifacts. As future work, we plan to experiment in different domains and investigate reliability, security, and efficiency aspects of the solution.

\begin{acks}

AI has been used to proofread the manuscript.
\end{acks}

%%
%% The next two lines define the bibliography style to be used, and
%% the bibliography file.
\bibliographystyle{ACM-Reference-Format}
\bibliography{refs}

\end{document}